\documentstyle[12pt,ssi,epsfig]{article}

\begin{document}
\title{Precision Measurement of The Anomalous Magnetic Moment of the Muon}
\author{C.S. \"{O}zben$^2$,~~~G.W. Bennett$^2$,~~B. Bousquet$^9$,~~H.N. Brown$^2$,~~G.M. Bunce$^2$,\\
R.M. Carey$^1$,~~~P. Cushman$^{9}$,~~~G.T. Danby$^2$,~~~~P.T. Debevec$^7$,~~~M. Deile$^{11}$,\\
H. Deng$^{11}$,~~~W. Deninger$^7$,~~~S.K. Dhawan$^{11}$,~~~V.P. Druzhinin$^3$,~~L. Duong$^{9}$,\\
E. Efstathiadis$^1$~~F.J.M. Farley$^{11}$, G.V. Fedotovich$^3$,~~S. Giron$^{9}$,~~F.E. Gray$^7$,\\
D. Grigoriev$^3$,~~~~~M. Grosse-Perdekamp$^{11}$,~~~~~A. Grossmann$^6$,~~~~M.F. Hare$^1$,\\
D.W. Hertzog$^7$,~~X. Huang$^1$,~~V.W. Hughes$^{11}$,~~M. Iwasaki$^{10}$,~~K. Jungmann$^5$,\\
D. Kawall$^{11}$,~~B.I. Khazin$^3$, J. Kindem$^{9}$, F. Krienen$^1$, I. Kronkvist$^{9}$, A. Lam$^1$,\\
R. Larsen$^2$,~~~Y.Y. Lee$^2$,~~~I. Logashenko$^1$,~~R. McNabb$^{9}$,~~W. Meng$^2$,~~J. Mi$^2$,\\
J.P. Miller$^1$,~~~W.M. Morse$^2$,~~~~D. Nikas$^2$,~~~~C.J.G. Onderwater$^7$,~~~Y. Orlov$^4$,\\
J.M. Paley$^1$,~~Q. Peng$^1$,~~~C.C. Polly$^7$,~~~J. Pretz$^{11}$,~~R. Prigl$^2$,~~G. zu Putlitz$^6$,\\
T. Qian$^9$,~~~~~S.I. Redin$^{3,11}$,~~~~~O. Rind$^1$,~~~~~B.L. Roberts$^1$,~~~~N.M. Ryskulov$^3$,\\
P. Shagin$^9$,~~~~~Y.K. Semertzidis$^2$,~~~~~Yu.M. Shatunov$^3$,~~~~E.P. Sichtermann$^{11}$,\\
E. Solodov$^3$,~~~M. Sossong$^7$,~~~~A. Steinmetz$^{11}$,~~~~L.R. Sulak$^1$,~~~~A. Trofimov$^1$,\\
D. Urner$^7$,~~~P. von Walter$^6$,~~~D. Warburton$^2$~~and~~A. Yamamoto$^8$.~~~~~~~~~~~~~~\vspace*{0.51cm} \\
\vspace*{1cm}
%\footnotesize{
{\it $^1$Boston University, Boston, Massachusetts 02215, USA} \\
{\it $^2$Brookhaven National Laboratory, Physics Dept., Upton, NY 11973, USA}\\
{\it $^3$Budker Institute of Nuclear Physics, Novosibirsk, Russia}\\
{\it $^4$Newman Laboratory, Cornell University, Ithaca, NY 14853, USA}\\
{\it $^5$Kernfysich Versneller Instituut, Rijksuniversiteit Groningen, Netherlands\\
{\it $^6$University of Heidelberg, Heidelberg 69120, Germany}\\
{\it $^7$University of Illinois, Physics Dept., Urbana-Champaign, IL 61801, USA}\\
{\it $^8$KEK, High Energy Acc. Res. Org., Tsukuba, Ibaraki 305-0801, Japan}\\
{\it $^{9}$University of Minnesota, Physics Dept., Minneapolis, MN 55455, USA}\\
{\it $^{10}$Tokyo Institute of Technology, Tokyo, Japan}\\
{\it $^{11}$Yale University, Physics Dept., New Haven, CT 06511,USA}
}}

\maketitle

\begin{abstract}
The muon g-2 experiment at Brookhaven National Laboratory measures the
anomalous magnetic moment of the muon, $a_\mu$, very precisely.  This
measurement tests the Standard Model.
The analysis for the data collected in 2000 (a $\mu^+$ run) is completed
and the accuracy on $a_\mu$ is improved to 0.7 ppm, including
statistical and systematic errors.
The data analysis was performed blindly between the precession
frequency and the field analysis groups in order to prevent a possible
bias in the $a_\mu$ result.   The result is
$a_{\mu}({\rm exp}) = 11~659~204(7)(5) \times 10^{-10}$ 
(0.7 ppm).  This paper features a detailed description of one of the 
four analyses and an update of the theory.
\end{abstract}

\section{Introduction}

The gyromagnetic ratio (g-factor) of a particle is defined as the ratio of
its magnetic moment to its spin angular momentum.
For a point-like spin-1/2 particle, the g value is predicted by the Dirac
equation to be equal to 2.  On the other hand, from experiments on
the hyperfine structure of hydrogen in the late 1940's, it was found that the g value of
an electron was not exactly 2 ($\approx$ 2.002).  In the first order this deviation is
due to the creation and absorption of a virtual photon by the particle.  This process
is described by quantum electrodynamics (QED) and the
electron g-2 experiments were precise tests of QED.  The Standard Model (SM)
calculations for the anomalous magnetic moment of the muon includes also the small contributions
from hadronic and weak interactions, in addition to  electromagnetic
(QED) interactions.  The QED contribution to the SM is the largest,
however it is the most precisely known.  On the other hand, the largest contribution to
the $a_\mu$ uncertainty comes from the hadronic interactions.
%The SM prediction of anomalous magnetic moment of muon \cite{amutheory} is
%a$_\mu=11~659~177(7)\times10^{-10}$(0.6 ppm).\\

\section{Experimental Method and Apparatus}

When a positive muon decays into a positron and two neutrinos,
parity is maximally violated so that the average positron
momentum is along the muon spin direction (in the muon rest frame).
Since observing the decay energy in the laboratory frame is equivalent 
to observing the decay angle in the center-of-mass system, 
the muon spin direction can be measured by counting the decay 
positrons above a certain energy threshold.  As a result,
the positron counting rate is modulated with the muon spin 
precession.  This is the main concept of the muon 
g-2 experiments\cite{g1}.\\

\noindent
In our experiment, polarized muons are injected into a superconducting
storage ring.  The ring provides a very uniform static dipole
magnetic field, $\vec{B}$. If the g-factor of a muon were exactly equal to
two, the spins and the momenta of the muons would stay parallel
during the storage.  This would mean the cyclotron frequency, $\omega_c$, of
the muon is equal to the spin precession frequency, $\omega_a$.
However, due to the anomaly on the magnetic moment, the spin precesses
faster than the momentum, $\omega_s=\omega_c(1+a_\mu\gamma$).
The muon spin precession angular frequency relative to
its momentum in the presence of a vertical focusing
electric field $\vec{E}$ and magnetic field is given by: 

\begin{equation}
\vec{\omega}_a = - \frac {e} {mc} \left[a_\mu\vec{B}-\left(a_\mu-
\frac {1} {\gamma^2-1} \right ) \vec{\beta}\times\vec{E} \right ]
\end{equation}

\noindent
where $a_\mu$ is the muon anomalous magnetic moment and $\gamma$ is
the relativistic Lorentz factor.\\

\noindent
The influence of the electric field on $\omega_a$ is eliminated by
using muons with the ``magic'' momentum $\gamma_\mu = 29.3$,
for which  $P = 3.09$~GeV/$c$.  The measurement of the angular
precession frequency $\omega_a$ and static field $B$ 
measured in terms of the proton NMR frequency, $\omega_p$,
determines $a_\mu$ as follows;
\begin{equation}
a_\mu =\frac {\frac {\omega_a} {\omega_p}} {\lambda-\frac {\omega_a} {\omega_p}}
\end{equation}

\noindent
where $\lambda=~\mu_\mu / \mu_p~=~3.183~345~39(10)$ \cite{lambda}.\\

\noindent
The counting rate $N(t)$ of decay positrons with energies above an energy threshold $E$
is modulated by the spin precession of the muon, ideally leading to 

\begin{equation}
N(t)~=~N_0(E)~e^{-t/{\tau_\mu}} \left( 1 + A {\mbox{cos}} \left[\omega_a t + \phi \right] \right)
\end{equation}

\noindent
where $N_0$ is the normalization, $\tau_\mu$ is the time-dilated muon lifetime,
$A$ is the energy-dependent asymmetry and $\phi$ is the phase. The
angular precession frequency $\omega_a$ is determined from a fit to
the experimental data.  Figure 1 shows the g-2
precession frequency data collected in 2000.  The error bars are in blue.
Figure 2 shows the comparison of the energy spectrum at the peak
and the trough of the g-2 cycles.  For a certain energy threshold,
the difference between two energy spectra is reflected in the asymmetry 
parameter we measure.  The energy dependence of the asymmetry can be
easily seen.  To minimize the statistical error on the precession
frequency, the energy threshold is chosen between 1.8-2.0 GeV.
For this threshold, the number of events times the square of 
the asymmetry is maximum.\\

\hspace*{-.7cm}
\begin{minipage}{0.50\linewidth}
\epsfig{file=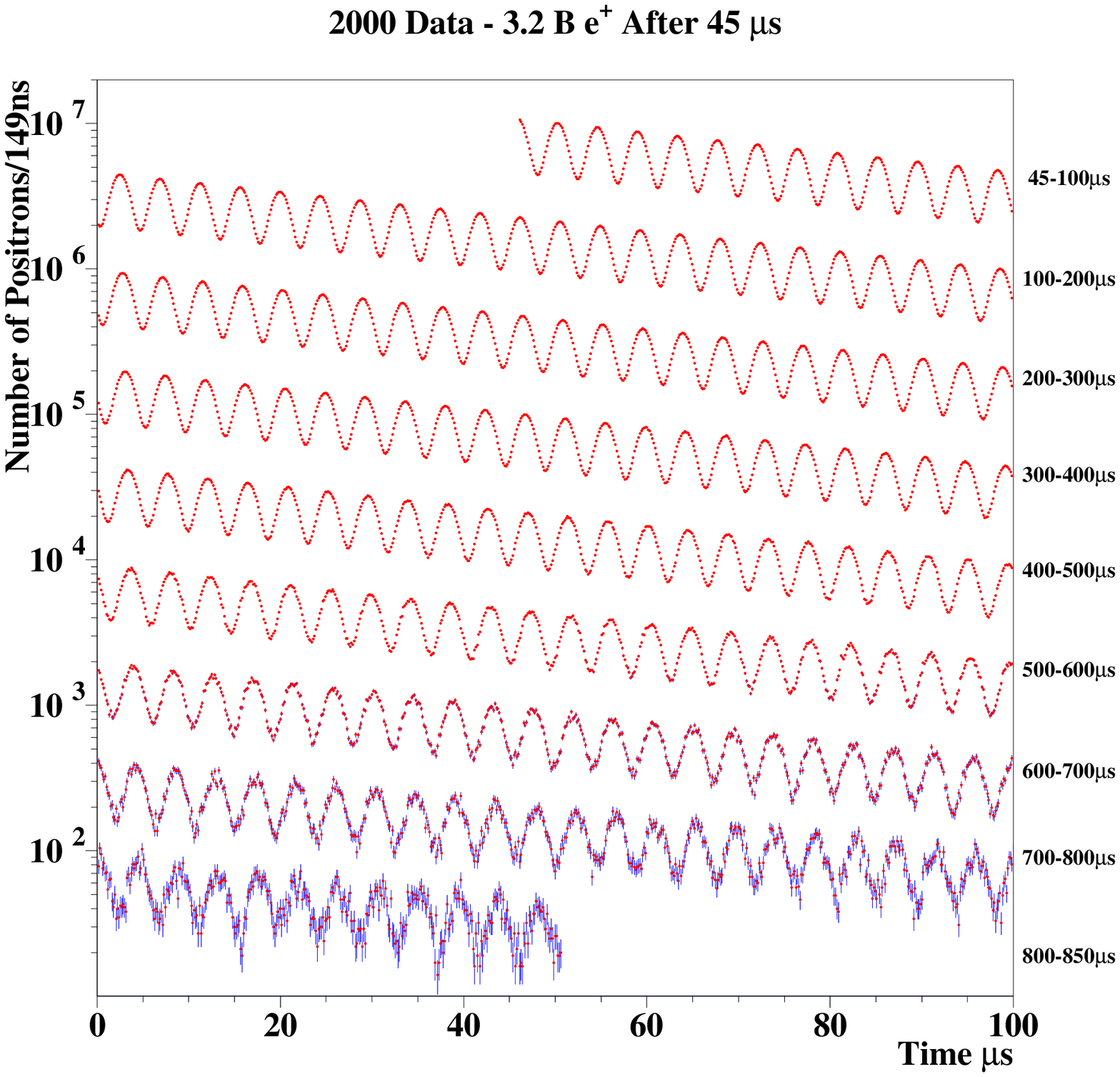, width=7cm, height=7cm}
{\bf Fig. 1 :} g-2 time spectrum.  The period of the g-2 is 4.36 $\mu$s and the dilated lifetime is 64.4 $\mu$s.
\end{minipage}
\hspace*{0.4cm}
\begin{minipage}{0.50\linewidth}
\epsfig{file=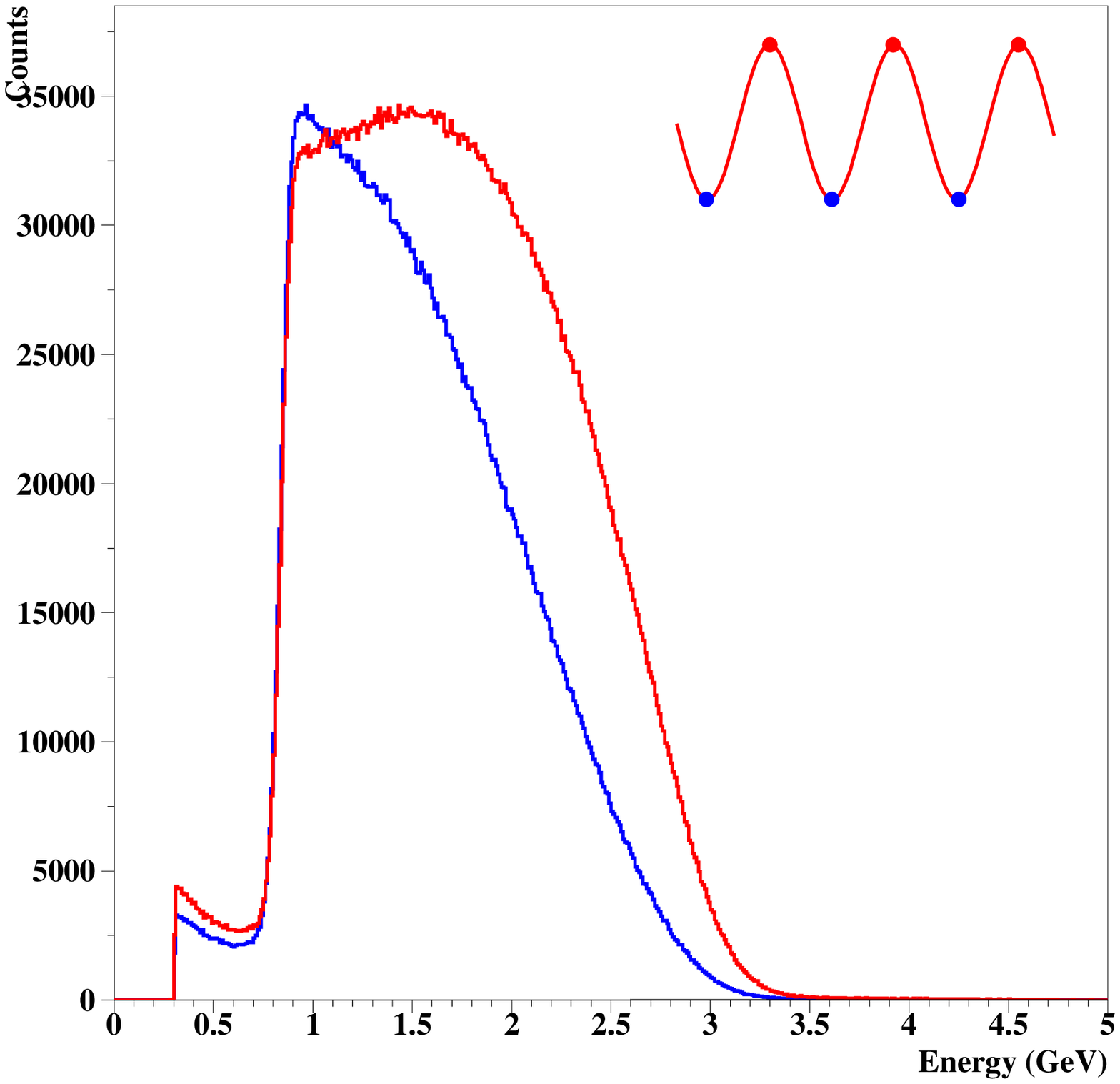, width=7cm, height=7cm}
{\bf Fig. 2 :} A comparison of the posit-\\
ron energy spectrum obtained at the\\
peak and at the trough of g-2. 
\end{minipage}\hfill  \\
\newline
\newline
\noindent
The BNL g-2 experiment uses a polarized muon beam produced by the
Alternating Gradient Synchrotron (AGS), which is the brightest proton
source in its energy class in the world.  Protons with an energy of 24 GeV hit a nickel
target; pions from this reaction are directed into a pion
decay channel where they decay into muons.  Muons are naturally
polarized when a small forward momentum bite is taken.  \\

\noindent
Our storage ring is a single superconducting C-shape magnet,
supplying a~1.45 T uniform field \cite{magnet}.
Beam muons are injected into the
storage ring through a superconducting inflector magnet, which locally cancels
the storage ring field and delivers muons
approximately parallel to the central orbit \cite{inflector}.  \\

\noindent
The BNL g-2 experiment started first data taking in 1997 using
pion injection into the storage ring \cite{1997run}.
This was similar to what was done in the last CERN experiment \cite{g1}.
Since the pions fill most of the phase space, there are always
a few daughter muons from pion decay which would be captured into
stable equilibrium orbits in the storage ring.  The disadvantage
of pion injection, was the high background level due to the pions themselves.
From 1998 on, our experiment used muon injection.  In order to put
the muons onto their equilibrium orbit, a fast magnetic 
kicker was used.  The background level was reduced dramatically and the number of
events was substantially increased with muon injection compared to
pion injection.  The kicker is located at 90$^0$ with respect to
the inflector and consists of three sets of parallel plates each 1.7 m 
long, carrying 5200 A for a very short time (basewidth 400 ns) \cite{kicker}. \\

\noindent
Electrostatic quadruples \cite{quads} are used to vertically confine the muons.
Four quadrupole assemblies are located in the ring, each
pulsed with $\pm$24 kV and covering 43$\%$  of
the azimuth in the storage ring.  \\

\noindent
Inside the ring, the decay positrons
are observed for approximately 600 $\mu$s by 24 electromagnetic shower 
calorimeters \cite{detectors} consisting of 
scintillating fibers embedded in lead.  All positrons above a certain threshold 
are digitized individually with 400 MHz waveform digitizers and the digitized waveforms 
are stored for the analysis.

\section{$\omega_a$ Analysis}
The g-2 frequency $\omega_a$ is determined by fitting the time
spectrum of positrons after data selection.  There were four independent
analyses of the precession frequency from the 2000 data.  However, only one
of them  \cite{cenap_2000} is going to be described here.\\

\noindent
When statistics are high, the influence of small effects becomes observable
in the data.  Deviations in the $\chi^2$ and the parameter stability
when experimental conditions are varied are the signature of the size of 
these effects.  The 2000 data amount to four times the number of events recorded in
1999.
%For that reason we observed additional effects.  
Positron pileup, coherent betatron oscillations (CBO) and muon losses were 
already observed in 1999 and accounted for in the fitting function \cite{1999_publication}.  
The mismatch between the inflector and storage ring acceptances is one of the main
sources of CBO. 
CBO is also caused by a non-ideal kick to muons when they first enter 
the ring.  Therefore, they are not captured in their equilibrium orbit, but 
rather oscillate about it.  
As a result of these oscillations, the beam gets closer and further away from the detectors 
resulting in a modulation of the counting rate.  CBO can be easily seen in the residuals 
constructed from the difference between the data and the 
ideal 5-parameter fit (Eq.3 and Fig. 3).  The Fourier transform of the residuals is shown in Figure 4.

\hspace*{-.9cm}
\begin{minipage}{0.50\linewidth}
\epsfig{file=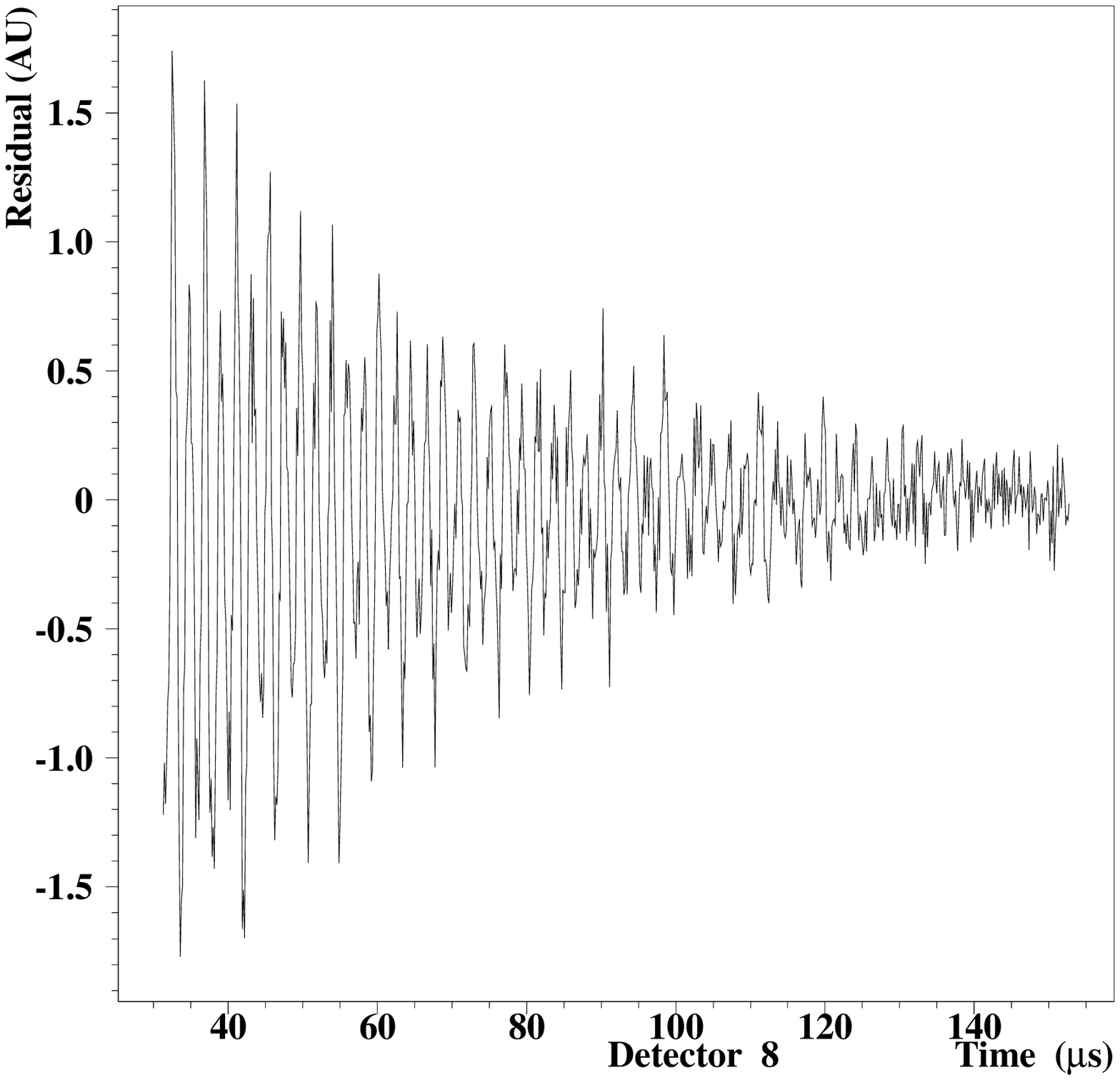, width=7cm, height=7cm}
\centerline{{\bf Fig. 3 :} Time spectrum of the residuals.}
\end{minipage}
\hspace*{0.4cm}
\begin{minipage}{0.50\linewidth}
\epsfig{file=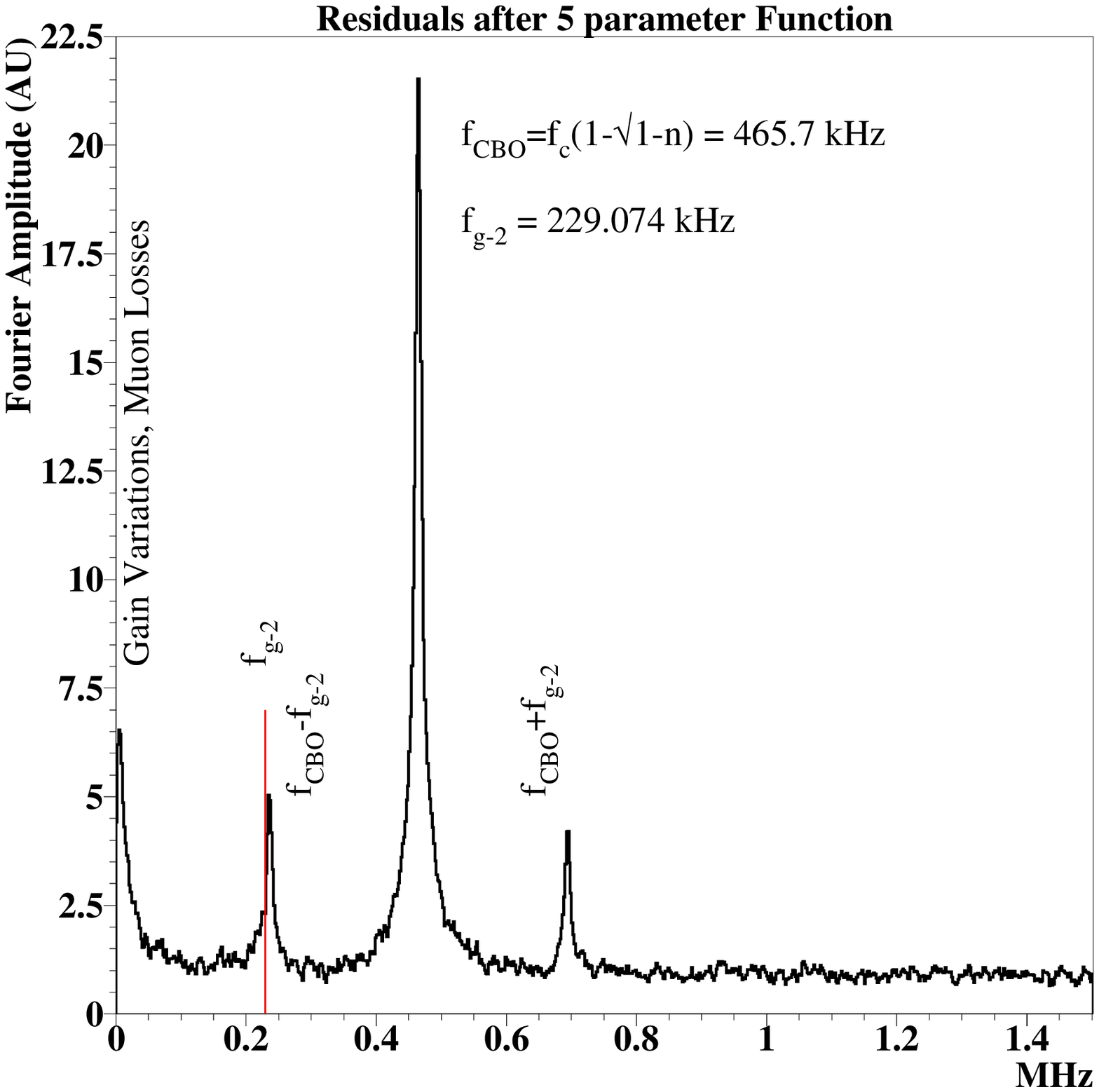, width=7cm, height=7cm}
\centerline{{\bf Fig. 4 :} Fourier spectrum of the residuals.}
\end{minipage}\hfill  \\

~\newline
\noindent
Modulations due to CBO are parameterized as:

\begin{equation}
F_{cbo}(t)=1+A_{cbo}~e^{-t/\tau_{cbo}}~{\mbox{cos}}\left[2\pi f_{cbo}t+\phi_{cbo}\right]
\end{equation}

\noindent
where $A_{cbo}$ is the amplitude of the modulation, $\tau_{cbo}$ is the coherence 
time of the damping, and $\phi_{cbo}$ is the phase.  The frequency $f_{cbo}$ is
fixed to the frequency determined by the Fourier analysis
(465.7$\pm$0.1 kHz) of the residuals.  With the larger data set from 2000, additionally
the g-2 asymmetry ($A$) and the phase ($\phi$) are also seen to be
modulated with the CBO.  Therefore, the fitting function (Eq. 3) was modified as follows :

\begin{equation}
N(t)~=~N_0~F_{cbo}(t)~e^{-t/\tau_\mu} \left( 1 + A(t) {\mbox{cos}} \left[\omega_a t + \phi(t) \right]
\right)
\end{equation}
where
\begin{equation}
A(t)~=~A~(~1~+~A_A~e^{-t/\tau_{cbo}}~{\mbox{cos}} \left[\omega_{cbo} t + \phi_A \right]~)
\end{equation}
and
\begin{equation}
\phi(t)~=~\phi~+~A_\phi~e^{-t/\tau_{cbo}}~{\mbox{cos}} \left[\omega_{cbo} t + \phi_\phi \right].
\end{equation}

\noindent
The amplitudes $A_{cbo}$, $A_A$ and $A_\phi$ are small ($\approx$ 1\%, $\approx$ 0.1\% and
$\approx$ 1 mrad, respectively) and are consistent with Monte Carlo simulation.\\

\noindent
To avoid beam resonances, the weak-focusing
index $n$ was set at 0.137, half way between two
neighboring resonances (Figure 5).  Running at a higher $n$ value was not
possible because the high voltage on the quads was limited to avoid breakdown.
Running at a lower $n$ value leads to storing less beam.
Running at $n~=~0.137$ made the CBO frequency close to two times
the g-2 frequency, which was a difficulty for the analysis, especially
for the fit.\\

\hspace*{-.9cm}
\begin{minipage}{0.50\linewidth}
\epsfig{file=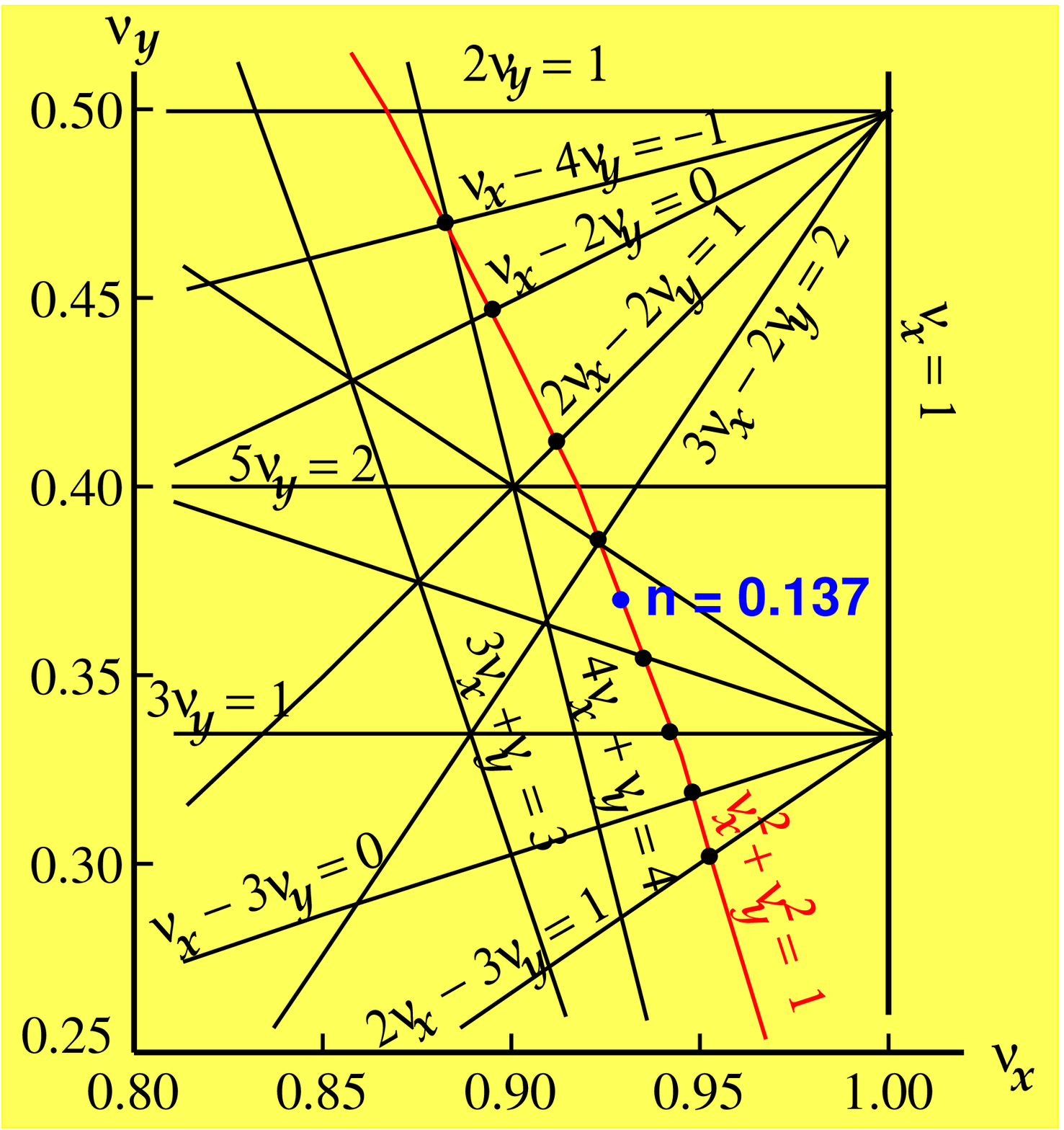, width=7cm, height=7cm}
{\bf Fig. 5 :} Beam tune and $n$ value. The lines represent the resonances.
\end{minipage}
\begin{minipage}{0.50\linewidth} ~\newline
\noindent
Pileup, the overlap of positron signals in time, leads to a
misidentification of positrons, their arrival time and their energy.
Pileup events have half the lifetime of the muons.  The size
of the effect is larger at early times compared to late times.
The way the data was stored permitted us to resolve pileup using the
data itself.  It was preferred to subtract the pileup events from
the data because including the pileup related terms into the \\
~\newline
\end{minipage} \hfill \vspace*{0.4cm}

\noindent
fitting function caused cross-talk between the fit parameters and 
an increase by a factor of two in the uncertainty of $\omega_a$.
Pileup pulses were reconstructed artificially from the data itself 
in the following way.  Every positron pulse above a relatively high 
threshold is digitized at 400 MHz for 
80 ns, which is called a ``WFD island''.  Positron pulses are 
fitted with a pulse-finding algorithm to determine energy and time
information.  This pulse-finding algorithm can resolve only events 
that are more than 3 ns apart.  After the main triggering 
pulse, there may be a second pulse on the WFD island, which 
carries the necessary energy and time information for artificial pileup 
construction.  When there is a main triggering pulse, we looked for a secondary 
pulse on the same WFD island within a time window offset 
from the trigger.  If there is a pulse, the main and the secondary pulses are added 
properly and the time is assigned from the energy-weighted 
time of the pair (Figure 6).  These constructed 
pileup events are then subtracted from the data to obtain 
a pileup-free time spectrum.  Figure 7 shows the fit to  
constructed pileup.\\

\hspace*{-.7cm}
\begin{minipage}{0.50\linewidth}
\epsfig{file=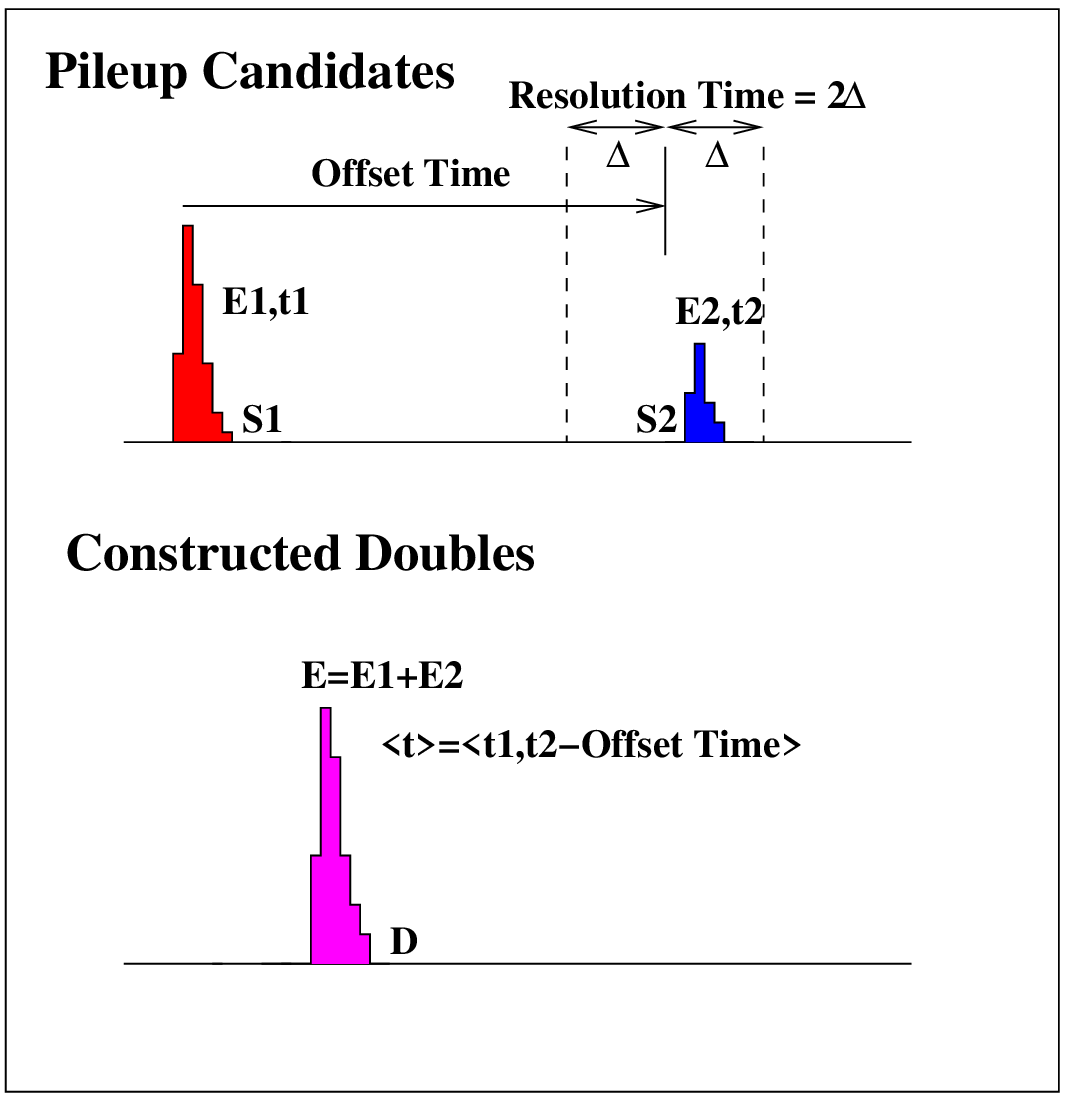, width=7cm, height=7cm}
{\bf Fig. 6 :} Schematic view of the pileup construction.
\end{minipage}
\hspace*{0.3cm}
\begin{minipage}{0.50\linewidth}
\epsfig{file=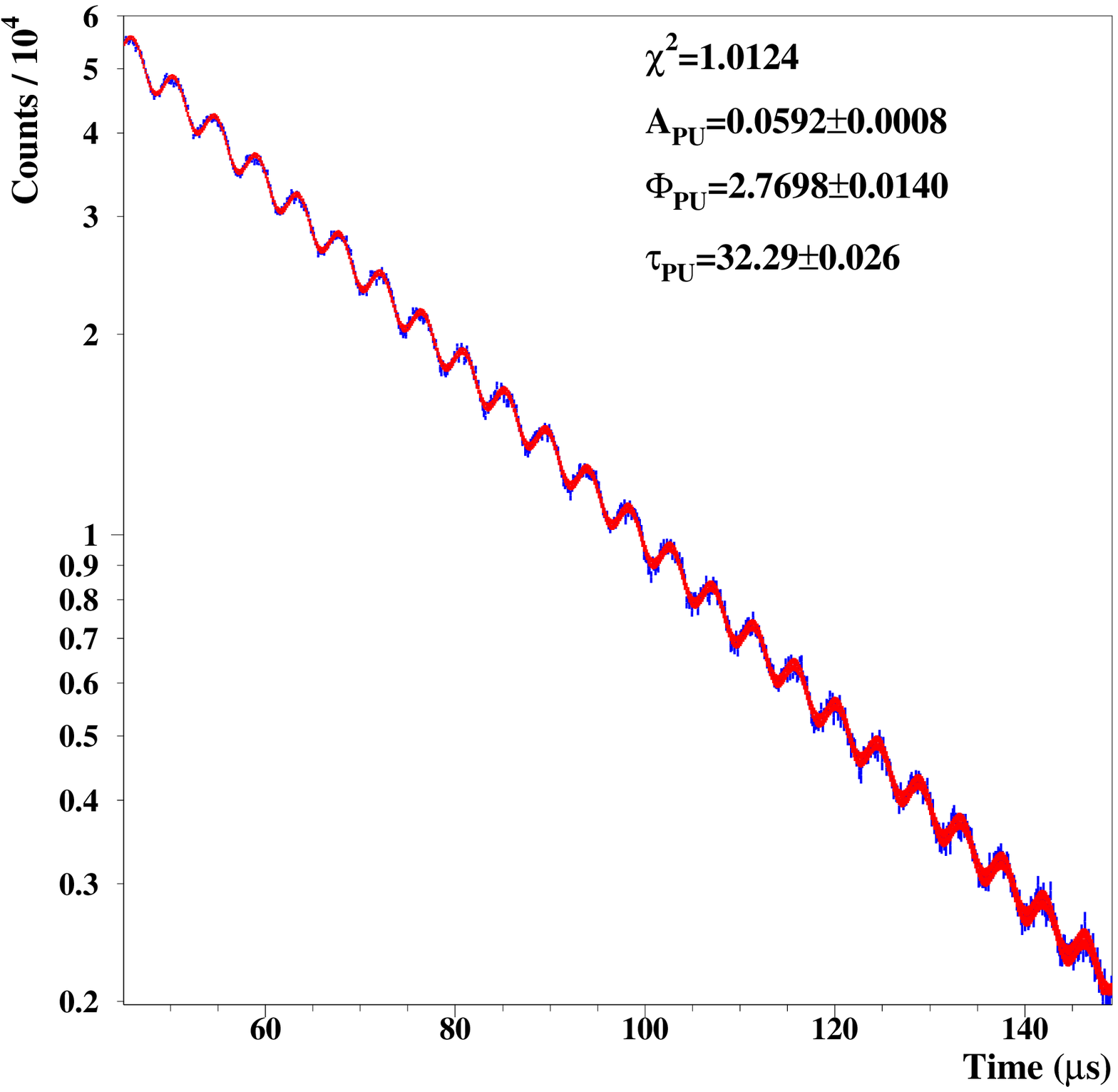, width=7.5cm, height=7.5cm}
\centerline{{\bf Fig. 7 :} Fit to constructed pileup.}
\end{minipage} \hfill   \\

\noindent
One of the tests of checking the pileup subtraction quality is to
look at the average energy versus time.  The average energy of
the detected positrons, for a given g-2 period, with and without the pileup 
subtraction, is shown in Figure 8.  The effect of pileup can be seen clearly 
in the upper curve.  After the pileup subtraction the average energy versus 
time is flat (lower curve).\\

\noindent
The next step in the analysis was to include muon losses.
The muon losses were determined two different ways.
The pileup subtracted data were fitted at later times ($\approx 300\mu$s) 
to an 8-parameter functional form (Ideal+CBO) and extrapolated to earlier times.
Only the CBO related parameters were determined at 50 $\mu$s since the 
lifetime of the CBO is $\approx 110 \mu$s.  
The ratio of the data to the extrapolated fitting function shows the
effect of the lost muons.
The second method is to look at three-fold 
coincidences through consecutive scintillator detectors.  Since energy
loss for muons is much smaller compared to positrons in the calorimeters, 
muons can travel between consecutive detectors
with little energy loss.  
%These events can be classified as lost muons.  
The dashed line in Figure 9 is the muon
losses determined from these three fold coincidences.  The agreement
between the two methods is good.  The muon loss time
spectrum is empirically added to the fitting function with a
scale factor, which is a fit parameter.  \\

\hspace*{-.7cm}
\begin{minipage}{0.50\linewidth}
\epsfig{file=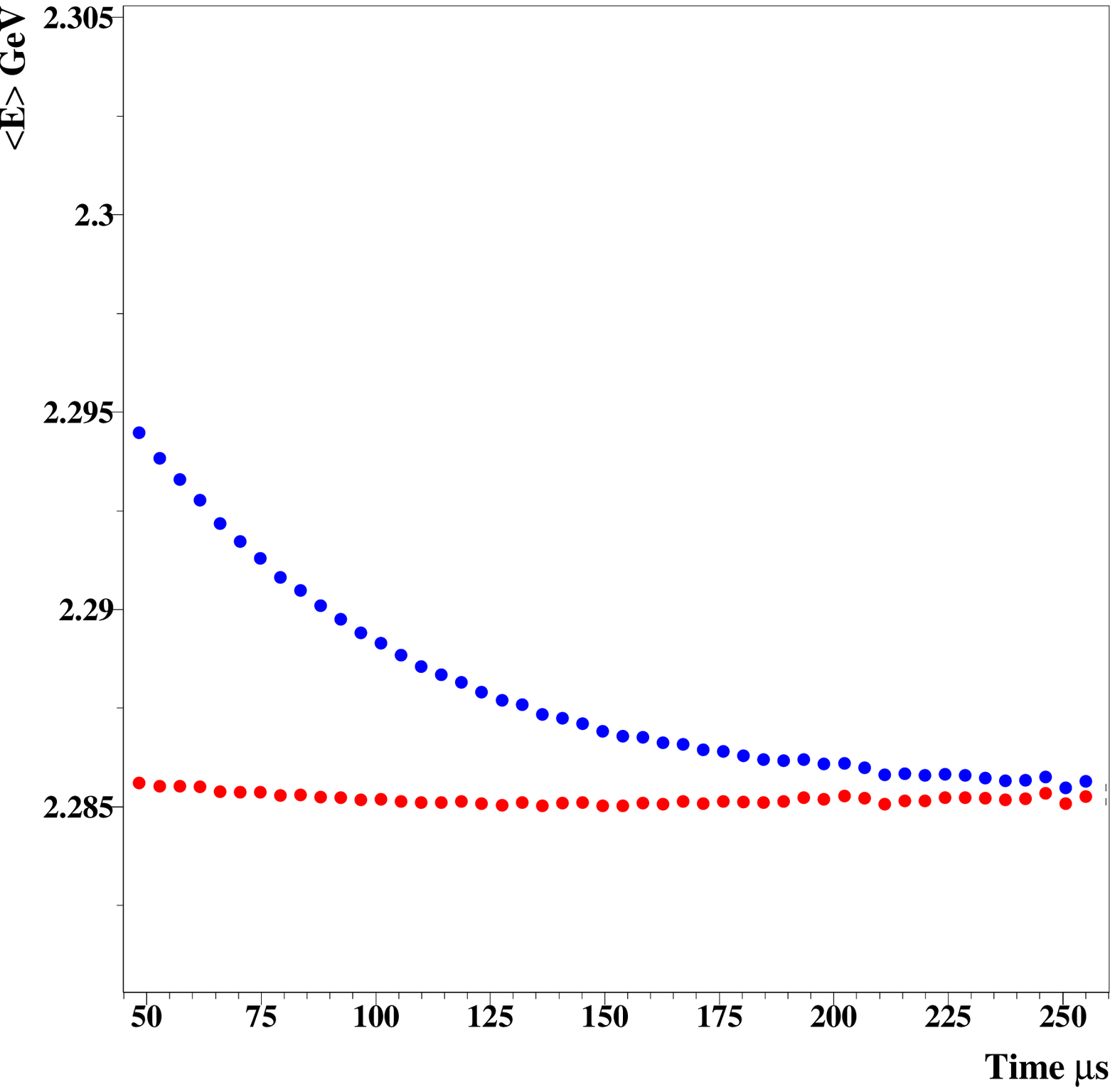, width=7cm, height=7cm}
{\bf Fig. 8 :} Average energy vs time with (lower) and without (upper) the pileup subtraction.
\end{minipage}
\hspace*{0.3cm}
\begin{minipage}{0.50\linewidth}
\epsfig{file=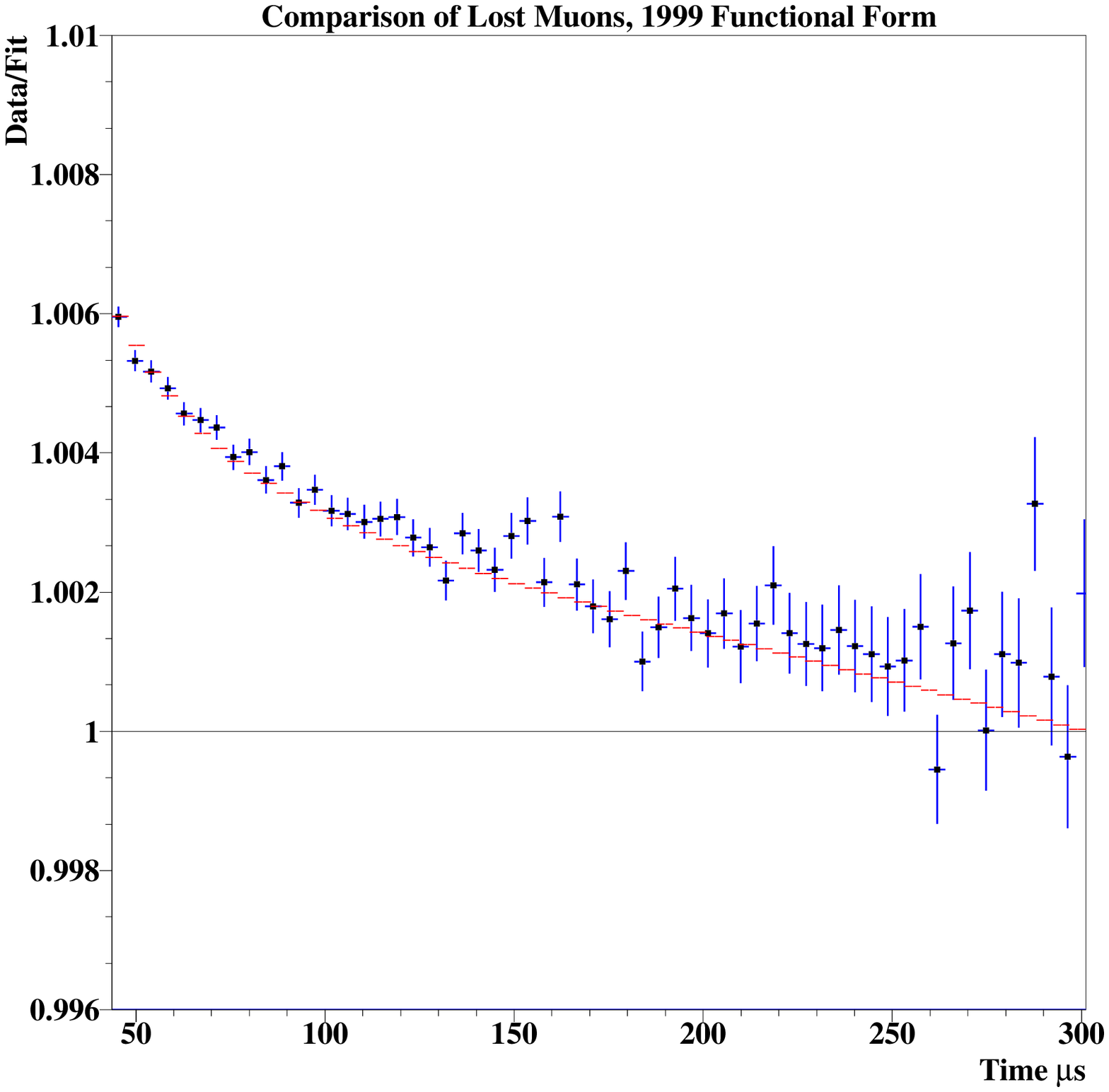, width=7cm, height=7cm}
\centerline{{\bf Fig. 9 :} Muon losses.}\\
~\newline
\end{minipage} \hfill \\

\noindent
It has been previously mentioned that the $n$ was 0.137.  This brought some difficulties
to the analysis since $\omega_{cbo}-\omega_a$ was very close
to $\omega_a$.  The fit had a difficult time to separate
these two frequencies.  Therefore, we had a
considerable systematic effect on $\omega_a$.  However, the CBO phase changes from 0 to $2\pi$
around the storage ring.  For that reason, when the data from individual
detectors are added, this effect becomes almost four times smaller.  On the other
hand when one looks at the precession frequency determined from the individual
detectors, the effect is visible.  Figure 10 shows the g-2 precession
frequency obtained from the individual detectors.  The fitting function
used here was the ideal five parameter function including modulation of the 
number of detected events caused by CBO (total 8 parameter).  
A fit to the data in Fig 10 gives a $\chi^2/DOF$ for
a fit to a constant, which is unacceptable.  When these data are fit to a sine 
wave (the CBO phase changes 0 to $2\pi$ around the ring), the
$\chi^2/DOF$ becomes acceptable (Fig. 10).  The central values on $\omega_a$ 
for both fits are very close. \\

\hspace*{-.7cm}
\begin{minipage}{0.50\linewidth}
\epsfig{file=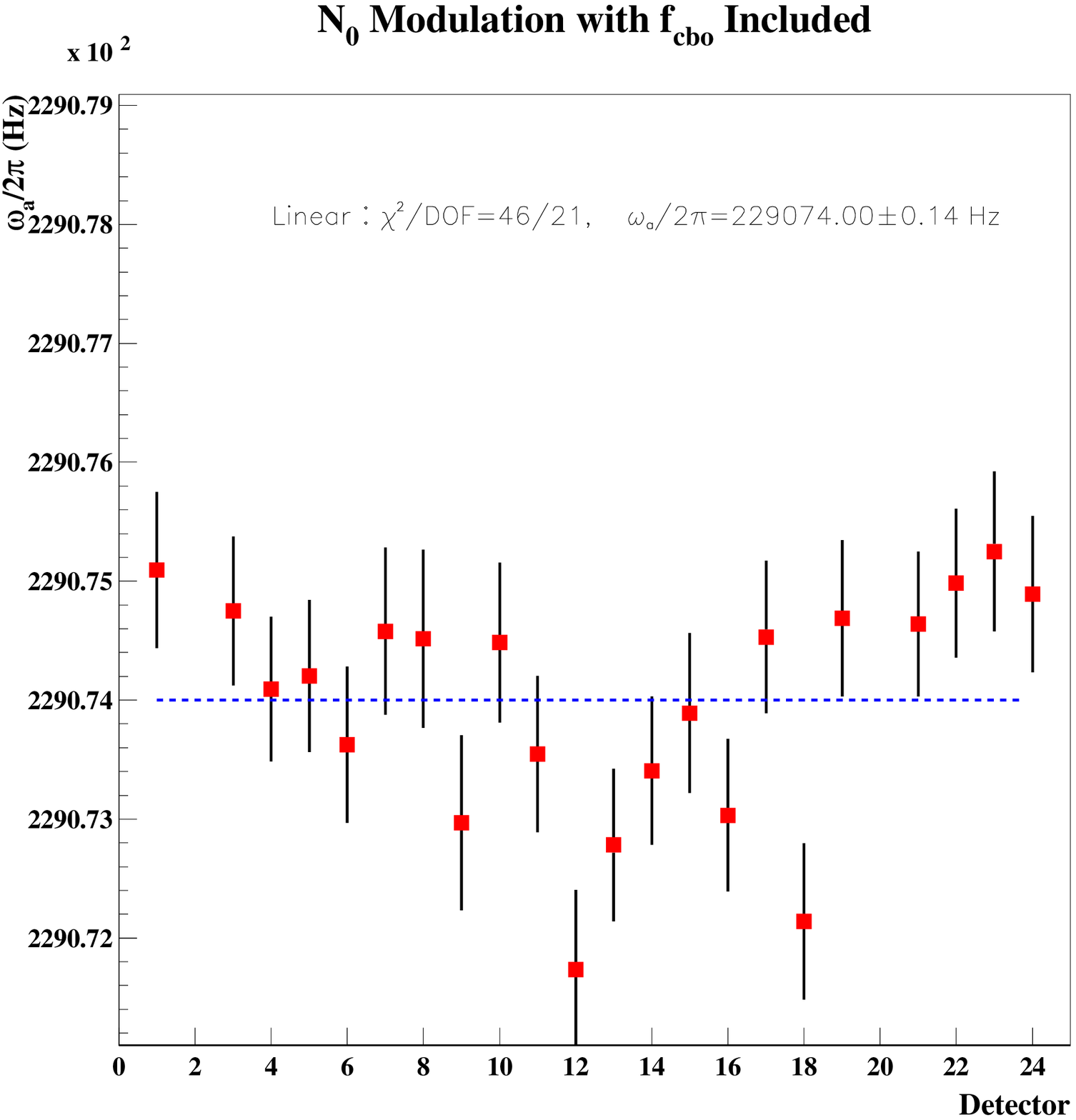, width=7cm, height=7cm}
\end{minipage}
\hspace*{0.3cm}
\begin{minipage}{0.50\linewidth}
\epsfig{file=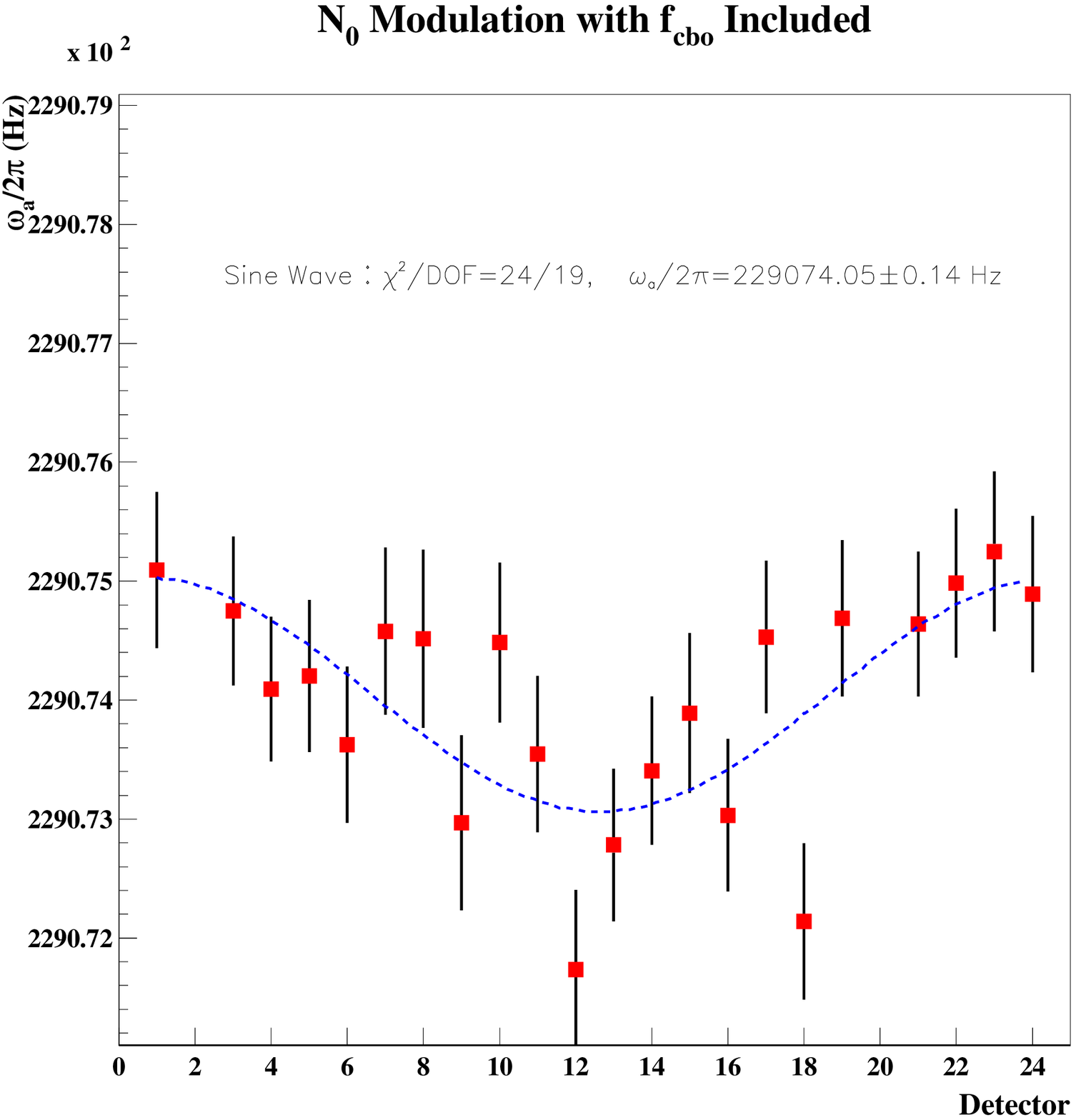, width=7cm, height=7cm}
\end{minipage} \hfill 
{\bf Fig. 10 :} Precession frequency vs detectors when the acceptance 
change due to CBO is included in the fits. \\

\noindent
The next step was to make the fitting function more precise by
adding the known effects of energy modulation.  One of these effects is 
the modulation of the g-2 asymmetry by CBO.  The amplitude of the sine wave
is reduced dramatically since most of the effect was removed (Fig. 11).
Figure 12 represents the result when both asymmetry and phase modulations
are included into the fit. \\

\hspace*{-.7cm}
\begin{minipage}{0.50\linewidth}
\epsfig{file=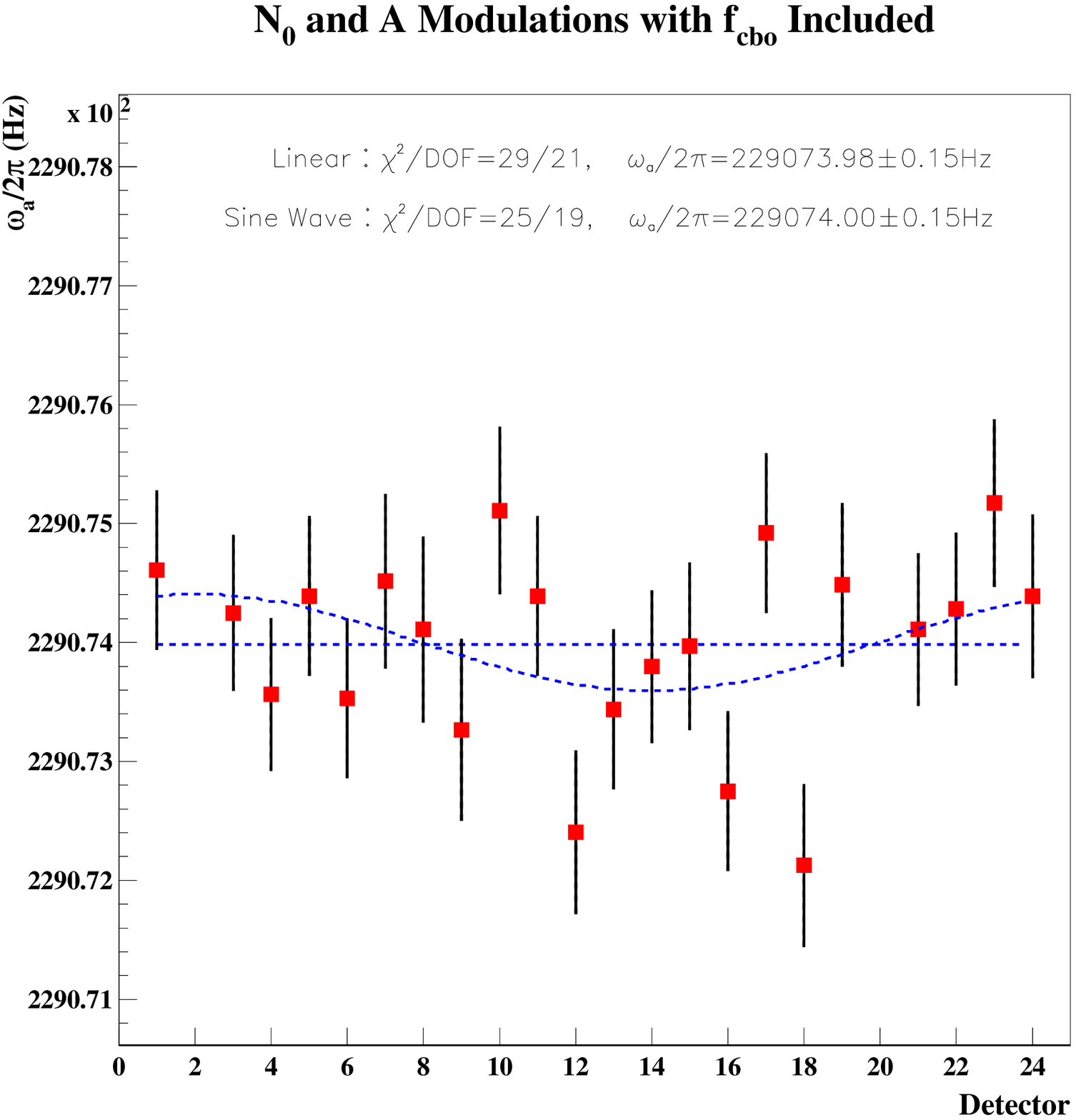, width=7cm, height=7cm}
{\bf Fig. 11 :} Precession frequency vs detectors when asymmetry
modulation due to CBO is included in the fit function.
\end{minipage}
\hspace*{0.3cm}
\begin{minipage}{0.50\linewidth}
\epsfig{file=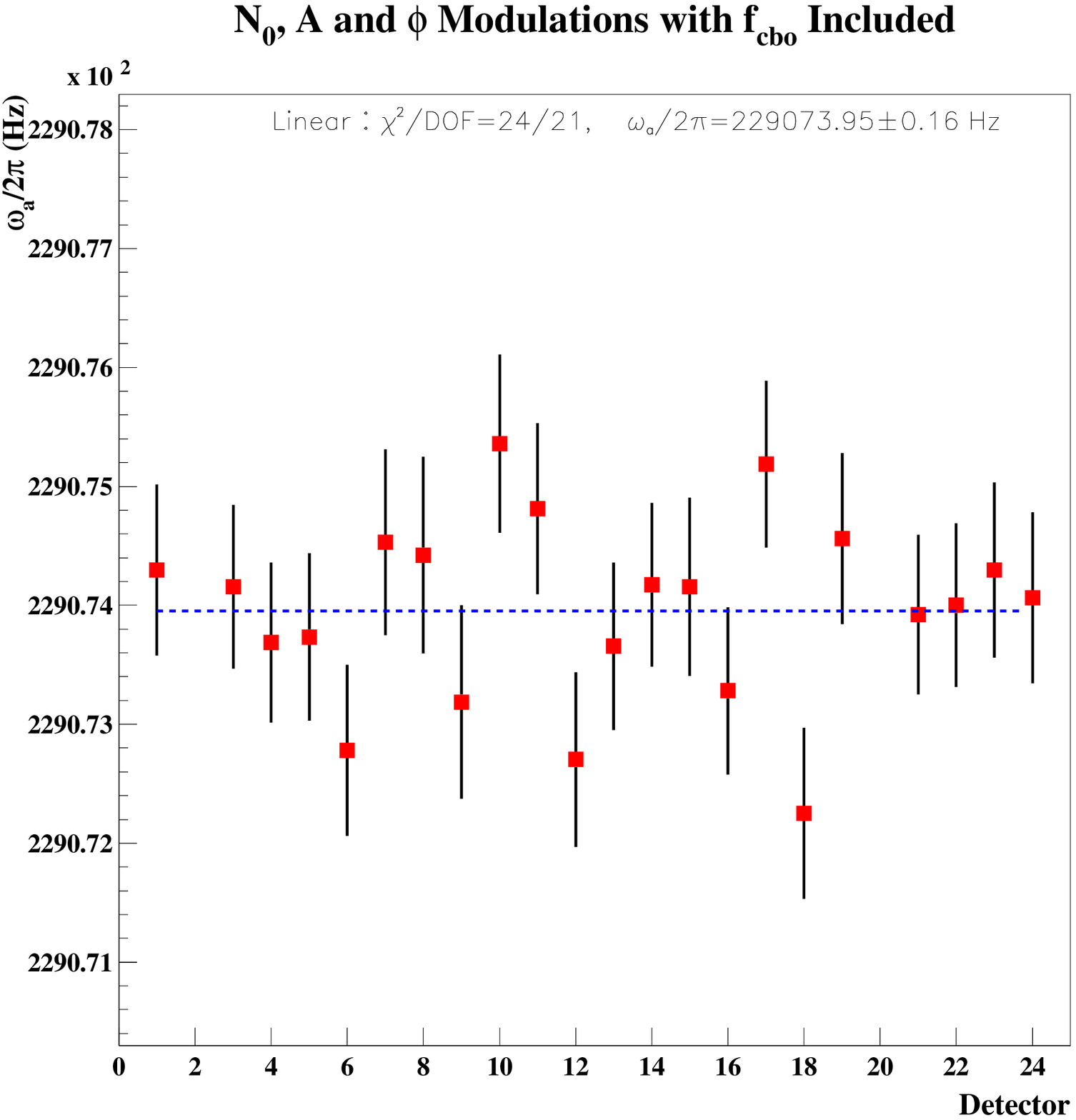, width=7cm, height=7cm}
{\bf Fig. 12 :} Precession frequency vs detectors when both asymmetry
and phase modulations are included in the fits.
\end{minipage} \hfill  \\

\noindent
This study showed that the effect can be removed completely
by adding all known effects to the fitting function.  However,
the center frequency value is not very sensitive to the type of
functional form used. \\

\noindent
Another method was pursued in the analysis.  That was to
sample the time spectrum with the CBO period \cite{Yuri}, so any CBO related effects can be
removed from the data and it can be fit to a five parameter ideal function.
The result of this method was consistent with the result of the 
method described in detail above.  \\

\noindent
The described analysis determined the precession frequency with 0.7 ppm statistical error.   
The most significant contributions to the systematic error were CBO (0.21 ppm), pileup (0.13 ppm), 
gain changes (0.13 ppm), lost muons (0.10 ppm) and fitting procedure (0.06 ppm). \\

\section{$\omega_p$ Analysis}
The magnetic field $B$ is obtained from NMR measurements of the proton
resonance frequency in water, which can be related to the free proton
resonance frequency $\omega_p$.\cite{NMR}  The field is continuously measured using about 150
fixed NMR probes distributed around the ring, in the top and bottom walls of the
vacuum chamber.  In addition to this measurement, the field inside the
storage ring where the muons are, is mapped with a trolley device.  
This device carries 17 NMR probes on it and the measurements were repeated 
periodically 2-3 times a week.  \\

\noindent
The trolley moves inside the storage ring and measures the field
around the ring without breaking the vacuum.  Figure 13 shows a two-dimensional
multipole expansion of the field averaged over azimuth from a trolley measurement.
The total systematic uncertainty from the field measurements is 0.24 ppm.  Figure 14 shows the
field map in the storage ring.\\

\begin{minipage}{0.90\linewidth}
\begin{center}
\epsfig{file=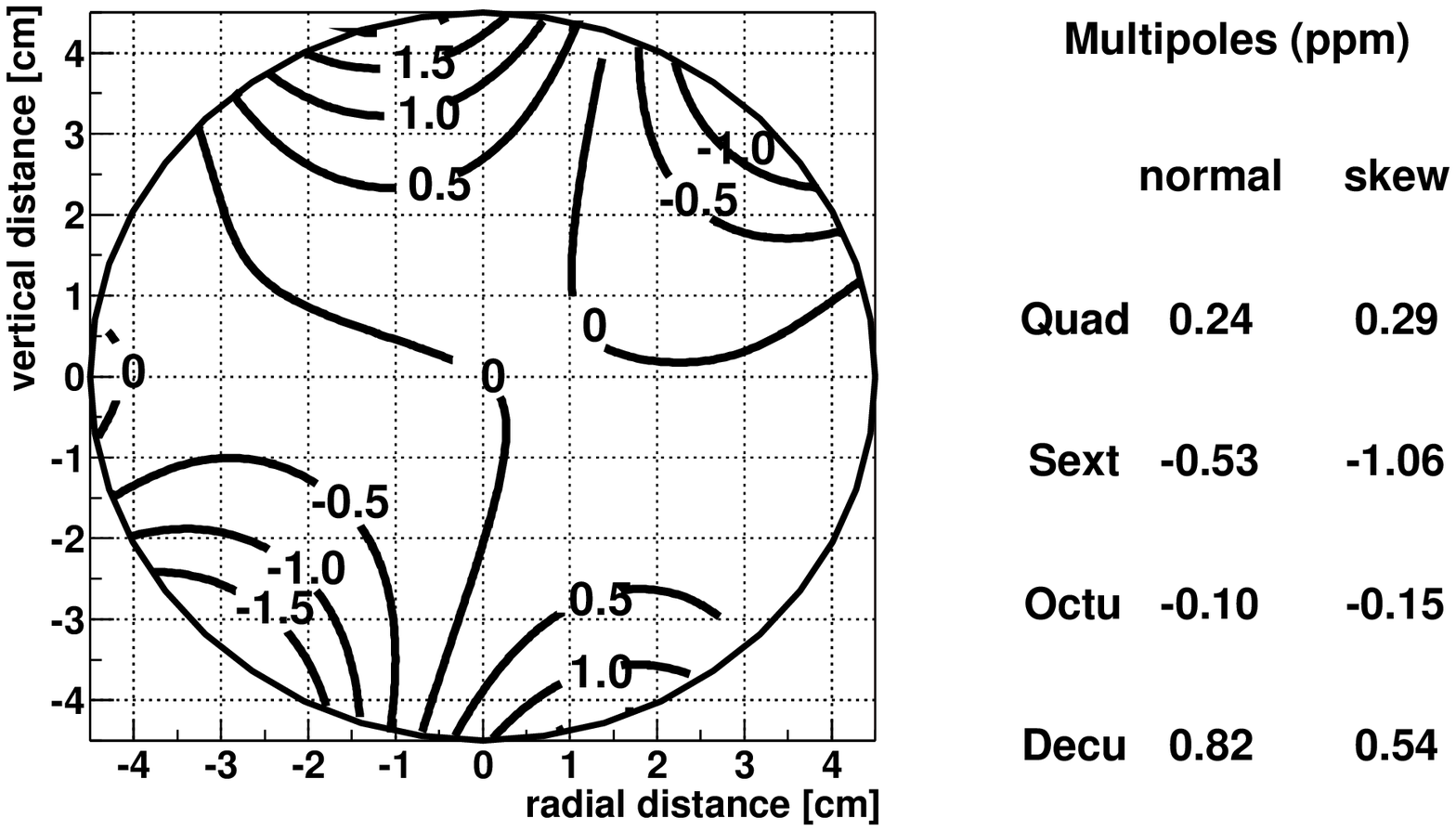, width=10cm, height=7cm}
\centerline{{\bf Fig. 13 :} Contour plot of multipole expansion.}
\end{center}
\end{minipage}

\begin{minipage}{0.90\linewidth}
\begin{center}
\epsfig{file=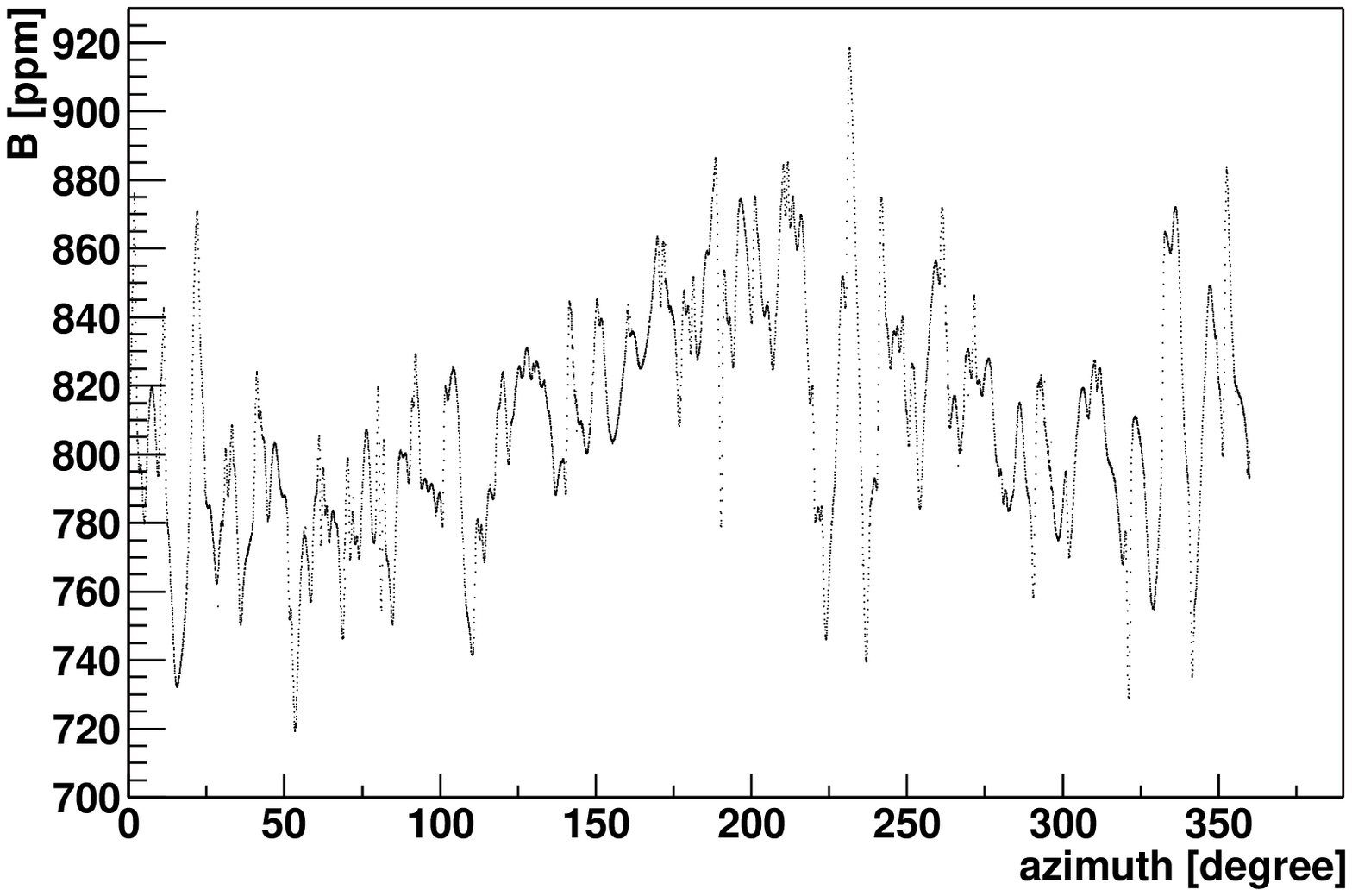, width=15cm, height=7cm}
\centerline{{\bf Fig. 14 :} Magnetic field map.}
\end{center}
\end{minipage} \\

\section{Result}
The anomaly $a_\mu$ can be obtained from the result of the independently analyzed
frequencies $\omega_a$ and $\omega_p$, and is determined to be

\begin{equation}
a_\mu=\frac{\omega_a} {\frac {e} {m\mu} <B>} =
\frac {\omega_a/\omega_p} {\mu_\mu/\mu_p-\omega_a/\omega_p}.
\end{equation}

\noindent
Figure 15 shows the comparison of the last three BNL results with
the SM evaluation.\\

\vspace*{-0.5cm}
\begin{minipage}{0.90\linewidth}
\begin{center}
\epsfig{file=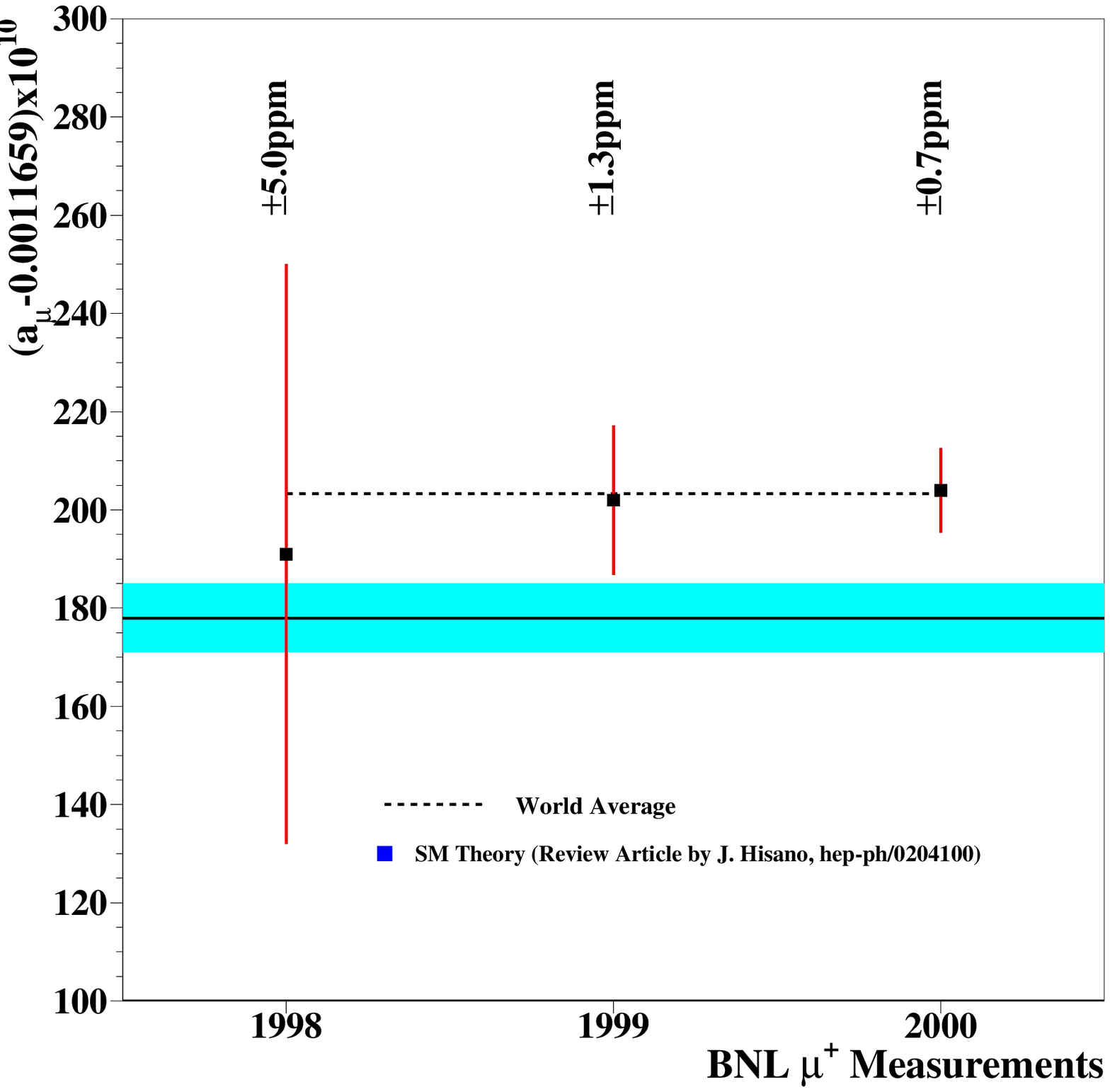, width=8.5cm, height=8.5cm}
\centerline{{\bf Fig. 15 :} Comparison of theory and recent BNL results.}
\end{center}
\end{minipage}\\ 

\noindent
The theoretical value of $a_{\mu}$ in the SM is determined
from $a_{\mu}({\rm SM}) = a_{\mu}({\rm QED})+ a_{\mu}({\rm had})+a_{\mu}({\rm weak})$. The
QED and weak contributions are given by\cite{13} $a_{\mu}({\rm QED})=
11\ 658\ 470.57(0.29)\times 10^{-10}$ (0.25 ppm)
and $a_{\mu}({\rm weak})=
15.1(04)\times 10^{-10}$ (0.03~ppm).
The leading-order contribution from
hadronic vacuum polarization contributes the largest uncertainty
to $a_{\mu}({\rm SM})$. Until
recently, $a_{\mu}({\rm had,1}) = 692(6) \times 10^{-10}$
(0.6 ppm) was the most reliable value,\cite{14,15} where data from both
hadronic $\tau$-decay and $e^+e^-$ annihilation were used to obtain
a single value for $a_{\mu}({\rm had,1})$.
Recently, two new
evaluations\cite{16,17} using the new $e^+e^-$ results from
Novosibirsk\cite{18}
have become available, and Ref. \cite{16} also employs data from hadronic
$\tau$-decay. While the two new analyses of $e^+e^-$ data agree quite well,
the value of $a_{\mu}({\rm had,1})$ obtained from $\tau$-decay does
not agree with the value obtained from $e^+e^-$ data.\cite{16}\\

\noindent
The higher-order
hadronic contributions include\cite{19}
$a_{\mu}({\rm had,2})= -10.0(0.6) \times 10^{-10}$ and the contribution
from hadronic light-by-light scattering is\cite{20}
$a_{\mu}({\rm had,lbl})= +8.6(3.2) \times 10^{-10}$.
Using the published value of $a_{\mu}({\rm had,1})$ from Ref. \cite{14}, the
standard model value is
$a_{\mu}({\rm SM}) = 11\ 659\ 177(7) \times 10^{-10}$ (0.6 ppm).\\

\noindent
From the most recent measurements at BNL, the muon anomalous magnetic
moment is determined as $a_{\mu}({\rm exp}) = 11~659~204(7)(5) \times 10^{-10}$ 
(0.7 ppm) \cite{bnl_2002} and the difference between $a_{\mu}({\rm exp})$ and 
$a_{\mu}({\rm SM})$ above is about 2.6 times the combined statistical and 
theoretical uncertainty.  If the new $e^+e^-$ evaluations are used\cite{16,17} 
the discrepancy is about 3 standard deviations, and using the $\tau$-analysis 
alone gives a 1.6 standard deviation discrepancy.\\

%\noindent
%The theoretical value of $a_\mu$ in the SM
%is determined as $a_\mu({\mbox{SM}})=a_\mu({\mbox{QED}})+a_\mu({\mbox{had}})+a_\mu({\mbox{weak}})$.  
%The size of the individual contributions are 
%$a_\mu({\mbox{QED}})=11~658~470.57(0.29)\times 10^{-10}$ (0.25 ppm) \cite{yedi}
%and $a_\mu({\mbox{weak}})=15.1(0.4)\times 10^{-10}$ (0.03 ppm) \cite{sekiz}.
%The leading contribution from hadronic part $a_\mu({\mbox{had,1}})$ is 
%$692(6)\times 10^{-10}$ (0.6 ppm) \cite{dokuz} has the largest
%uncertainty. There are other recent evaluations \cite{on,onbir} 
%on $a_\mu({\mbox{had,1}})$.  However, these evaluations are based on the 
%preliminary Novasibirsk data, which have been superseded \cite{oniki}
%and therefore they are not shown and not used for the comparison to
%our experimental result.  Higher-order contributions \cite{onuc} include 
%$a_\mu({\mbox{had,2}})=-10.0(0.6)\times 10^{-10}$ and
%the contribution from light-by-light scattering \cite{ondort},
%$a_\mu({\mbox{had,lbl}})=+8.6(3.2)\times 10^{-10}$.\cite{hlbl}
%Hence, the value of $a_\mu$(SM) is currently evaluated to be,
%$a_\mu({\mbox{SM}})=11~659~177(7) \times 10^{-10}$ (0.6 ppm).\\

%\noindent
%From the most recent measurements at BNL, the muon  
%anomalous magnetic moment is determined as
%$a_\mu({\mbox{exp}})=11~659~204(7)(5) \times 10^{-10}$ (0.7 ppm) \cite{bnl_2002}. 
%The difference of $a_\mu$(exp) and $a_\mu$(SM) is about 2.6 times 
%combined statistical and theoretical uncertainty.\\

\noindent
In the 2001 run, the muon g-2 experiment at BNL took data with negative muons
with a similar statistical power to the 2000 data.  This measurement will
provide a test of CPT violation and also an improved value of $a_\mu$.

\end{document}